\documentclass[twocolumn,showpacs,preprintnumbers]{revtex4-1}
\usepackage{amsfonts}
\usepackage{amsmath}

\usepackage{amssymb}
\usepackage{graphicx}
\usepackage{bm}

\newcommand{\ie}{{\it i.e.}}

\newcommand{\etal}{{\it et al.}}
\providecommand{\U}[1]{\protect\rule{.1in}{.1in}}

\begin{document}
\title{Quantum phase transition in an atom-molecule conversion  system with atomic hopping}
\author{Ning-Ju Hui${}^1$, Li-Hua Lu${}^1$*, Xiao-Qiang Xu${}^2$ and You-Quan Li${}^1$}
\affiliation{${}^1$Department of Physics, Zhejiang University, Hangzhou 310027, China\\
${}^2$ Department of Physics, Hangzhou Normal University, Hangzhou 310036, China}

\begin{abstract}
The quantum phase transition in an atom-molecule conversion system with atomic hopping between different hyperfine states is studied. In mean field approximation, we give the phase diagram whose phase boundary only depends on the atomic hopping strength and the atom-molecule energy detuning  but not on the atomic interaction. Such a phase boundary is further confirmed by the fidelity of the ground state and the energy gap between the first-excited state and the ground one. In comparison to mean field approximation, we also study the quantum phase transition in full quantum method, where the phase boundary can be affected by the particle number of the system. Whereas, with the help of finite-size scaling behaviors of energy gap, fidelity susceptibility and the first-order derivative of entanglement entropy, we show that one can obtain  the same phase boundary by the MFA and full quantum methods in the limit of $N\rightarrow \infty$. Additionally, our results show that the quantum phase transition can happen at the critical value of the atomic hopping strength even if the atom-molecule energy detuning is fixed on a certain value, which provides one a new way to control the quantum phase transition.

\end{abstract}
 \received{\today } \maketitle

\maketitle
\section{Introduction}
The quantum phase transition (QPT) describes an abrupt change in the ground state of a many-body system as some system parameters going across a critical point (at zero temperature). Quantum phase transitions (QPTs) in the systems of the quantum Hall, superconductor and ultracold atoms have been studied extensively \cite{QH, SC, QPT}. Ultracold atomic systems, especially Bose-Einstein condensates (BECs), provide us a good platform to study the QPT.
The experiment on ultracold atoms in an optical lattice by Bloch \etal~has given a good example of the QPT from a superfluid to a Mott insulator (SF-MI) \cite{SF-MI1}.
Then the SF-MI transition in the ultracold atomic systems was discussed widely~\cite{SF-MI2, SF-MI3}. Besides the SF-MI transition, the transition from non-entangled to entangled states in two-mode BECs ~\cite{QPT_Eentropy_Op}, from  degenerate and non-degenerate ground states in the extended boson Josephson-junction model~\cite{T-QPT},  and from a pure molecule state to a mixed atom-molecule one  in an atom-molecule model~\cite{entanglement, QPT_AM1, QPT_AM2, QPT_Fu, BA} have been investigated.  Note that in the above QPTs, one can control the QPT by changing different system parameters such as atom-pair tunneling strength and energy detuning between the atomic and the molecular states.

Molecular BECs are versatile in physical studies because, in comparison to atomic BECs, they have more freedoms to be controlled. For example,
the ultracold polar molecules have been used to study the ultracold chemistry, quantum many-body physics and quantum information science.  The polar molecules with dipole moment may be either heteronuclear molecules or  homonuclear molecules. The heteronuclear molecules with large electric-dipole moment have been investigated  both in theory and experiment widely~\cite{AHM2_Exp, AHM3_Exp, AHM2_The, AHM3_The}. The homonuclear molecules with large dipole moment, such as Rydberg molecules, have been produced by T. Pfau \etal~in experiments~\cite{RM1} and studied widely~\cite{RM2, RM3, RM4}. One Rydberg molecule consists of a single kind of atoms in different states, and the atoms can jump between the ground state and Rydberg states, which provides probability to study the QPT in atom-molecule conversion system with atomic hopping. Although  the quantum phase transition of the ground state has been studied in the atom-molecule system~\cite{entanglement, QPT_AM1, QPT_AM2, QPT_Fu, BA}, the atoms in that system were in the same state and the hopping
between different hyperfine atomic states was not considered.  So it is need to consider the atom-molecule conversion system with atomic hopping and study the effect of the hopping strength between different hyperfine atomic states on the quantum phase transition.

  In this paper, we study the quantum phase transition of a atom-molecule conversion system with atomic hopping between different hyperfine atomic states in both mean field approximation (MFA) and full quantum methods. It is interesting to find that the QPT can still appear in the system as the hopping strength increasing even if the atom-molecule energy detuning is fixed on a certain value, which is different from Ref.~\cite{QPT_Fu} where the QPT was induced by changing the energy detuning. In MFA, we show that the QPT exists in the thermodynamic limit (\ie, $N\rightarrow\infty$) and give the phase diagram whose phase boundary only depends on the atomic hopping strength and the atom-molecule energy detuning but not the atomic interaction.   In full quantum method, we characterize the QPT with the help of energy gap, fidelity susceptibility and the first-order derivative of entanglement entropy, which give the same phase boundary. We also show that one can obtain the same phase boundary by the MFA and full quantum methods in the limit of $N\rightarrow \infty$ through studying  the finite-size scaling behaviors of energy gap, fidelity susceptibility and the first-order derivative of entanglement entropy, which further confirm  the existence of the QPT in the atom-molecule conversion system with atomic hopping.

In the next section, we give the model of the system and the general phase diagram in mean field approximation. The energy gap and fidelity are also studied to characterize the QPT. In Sect.~\ref{sec:QPT}, we study the QPT from a pure molecule state to an atom-molecule mixed state in full quantum method. The energy  gap, fidelity susceptibility,  entanglement entropy,  first-order derivative of entanglement entropy, and their scaling behaviors are investigated.  In the last section, we give a brief summary.

\section{Model and Phase diagram}\label{sec:M-MFA}
We consider a three-component atom-to-molecule conversion system where the atoms can jump between two hyperfine atomic states.
\begin{eqnarray}
\label{model}
H&=&
-J(\hat{a}^{\dagger}_1\hat{a}_2+\hat{a}^{\dagger}_2\hat{a}_1)+\frac{\delta_{a}}{2}
(\hat{a}^{\dagger}_1\hat{a}_1-\hat{a}^{\dagger}_2\hat{a}_2)\nonumber\\
&&+\frac{U_a}{2}(\hat{a}^{\dagger}_1\hat{a}^{\dagger}_1\hat{a}_1\hat{a}_1
+\hat{a}^{\dagger}_2\hat{a}^{\dagger}_2\hat{a}_2\hat{a}_2)
+U^\prime_a \hat{a}^{\dagger}_1\hat{a}^{\dagger}_2\hat{a}_2\hat{a}_1\nonumber\\
&&+g^\prime(\hat{b}^{\dagger}\hat{a}_1\hat{a}_2+\hat{a}^{\dagger}_1\hat{a}^{\dagger}_2\hat{b})
-\frac{\delta_b}{2}(\hat{a}^{\dagger}_1\hat{a}_1+\hat{a}^{\dagger}_2\hat{a}_2-\hat{b}^{\dagger}\hat{b}),
\end{eqnarray}
where $\hat{a}_i$ and $\hat{a}^\dagger_i$ ($i=1,2$) denote that annihilate and create a atom in the $i$th hyperfine atomic states, and  $\hat{b}$ and $\hat{b}^\dagger$ denote that annihilate and create a molecule, respectively. Here $J$ refers to the hopping strength between the two atomic components,  $g^\prime$ the atom-molecule coupling strength, $U_a$ ($U_a^\prime$) the strength of atomic intracomponent (intercomponent) interaction, $\delta_{a}$ the energy detuning between the two atomic components, and $\delta_b$ the energy detuning between  atomic  and molecular components.

In order to study the property of the system we considered, we need to give the dynamical equations of the
 system in mean-field approximation (MFA). We know that with the help of Heisenberg motion equation for operators, one can easily give the evolution equations for operator $\hat{a}_i$ and $\hat{b}$. In mean-field approximation, the average value of the operators can be replaced by their average values, \ie, $\alpha_1=\langle\hat{a}_1\rangle/\sqrt{N}$, $\alpha_2=\langle\hat{a}_2\rangle/\sqrt{N}$ and $\beta=\langle\hat{b}\rangle/\sqrt{N
}$.  Then we can obtain the evolution equations for $\alpha_1$, $\alpha_2$ and $\beta$  with the help of the aforementioned evolution equations for $\hat{a}_i$ and $\hat{b}$. Here $\alpha_i$ and $\beta$ satisfy the conservation law $|\alpha_1|^2+|\alpha_2|^2+2|\beta|^2=1$ and $N$ is the total particle number.
For convenience to study the properties of fixed points, we assume $\alpha_{1}=\sqrt{\rho _{1}}e^{i\theta _{1}}$, $\alpha_{2}=\sqrt{\rho _{2}}e^{i\theta
_{2}}$ and $\beta=\sqrt{\rho _{b}}e^{i\theta_{b}}$, where $\rho_1+\rho_2+2\rho_b=1$.  The mean-field dynamical equations of the system can be written as
\begin{eqnarray}
\label{evo_cl}
\frac{d}{dt}\phi_{a}&=&[\frac{J\cos (2\phi _{a})-g\sqrt{\rho
_{b}}\cos\phi}{\sqrt{(1-2\rho _{b})^{2}-z^{2}}}+\frac{U}{2}-\frac{
U^{\prime }}{2}]z+\frac{\delta_{a}}{2},\nonumber\\
\frac{d}{dt}\phi~&=&[\frac{2J\cos (2\phi _{a})-2g\sqrt{\rho
_{b}}\cos\phi}{\sqrt{(1-2\rho _{b})^{2}-z^{2}}}-U-
U^{\prime }](1-2\rho _{b})\nonumber\\
&&+\frac{g}{2}\sqrt{\frac{(1-2\rho _{b})^{2}-z^{2}}{\rho _{b}}
}\cos\phi+\frac{3\delta_b}{2},\nonumber\\
\frac{d}{dt}z~&=&-2J\sqrt{(1-2\rho
_{b})^{2}-z^{2}}\sin (2\phi _{a}),\nonumber\\
\frac{d}{dt}\rho _{b} &=&g\sqrt{
((1-2\rho _{b})^{2}-z^{2})\rho _{b}}\sin\phi,
\end{eqnarray}
where $\phi _{a}=(\theta _{2}-\theta _{1})/2$, $\phi=\theta _{1}+\theta _{2}-\theta
_{b}$, $z=\rho _{1}-\rho _{2}$. Here  the renormalized  parameters are $U=N U_a$, $U^{\prime }=N U^\prime_a$ and $g=\sqrt{N}g^\prime$.  The equations~(\ref{evo_cl}) satisfy the Hamiltonian canonical equations, \ie, $\frac{d}{dt}\phi _{a}=\frac{\partial H_{\texttt{MF}}}{\partial z}$, $\frac{d}{dt}z=-\frac{\partial H_{\texttt{MF}}}{\partial \phi _{a}}$, $\frac{d}{dt}\phi=\frac{\partial H_{\texttt{MF}}}{\partial \rho _{b}}$ and $\frac{d}{dt}\rho _{b}=-\frac{\partial H_{\texttt{MF}}}{\partial \phi }$, where the Hamiltonian in MFA reads
\begin{eqnarray}
\label{hami_cl}
H_{\texttt{MF}}&=&-J\sqrt{(1-2\rho _{b})^{2}-z^{2}}\cos (2\phi _{a})+\frac{\delta_{a}}{2}z \nonumber\\
&&+\frac{U}{4}
((1-2\rho _{b})^{2}+z^{2})+\frac{U^{\prime }}{4}((1-2\rho
_{b})^{2}-z^{2}) \nonumber\\
&&+g\sqrt{[(1-2\rho _{b})^{2}-z^{2}]\rho _{b}}\cos\phi+\frac{\delta_b}{2}(3\rho
_{b}-1).\nonumber\\
\end{eqnarray}

\begin{figure}[tbph]
\includegraphics[width=70mm]{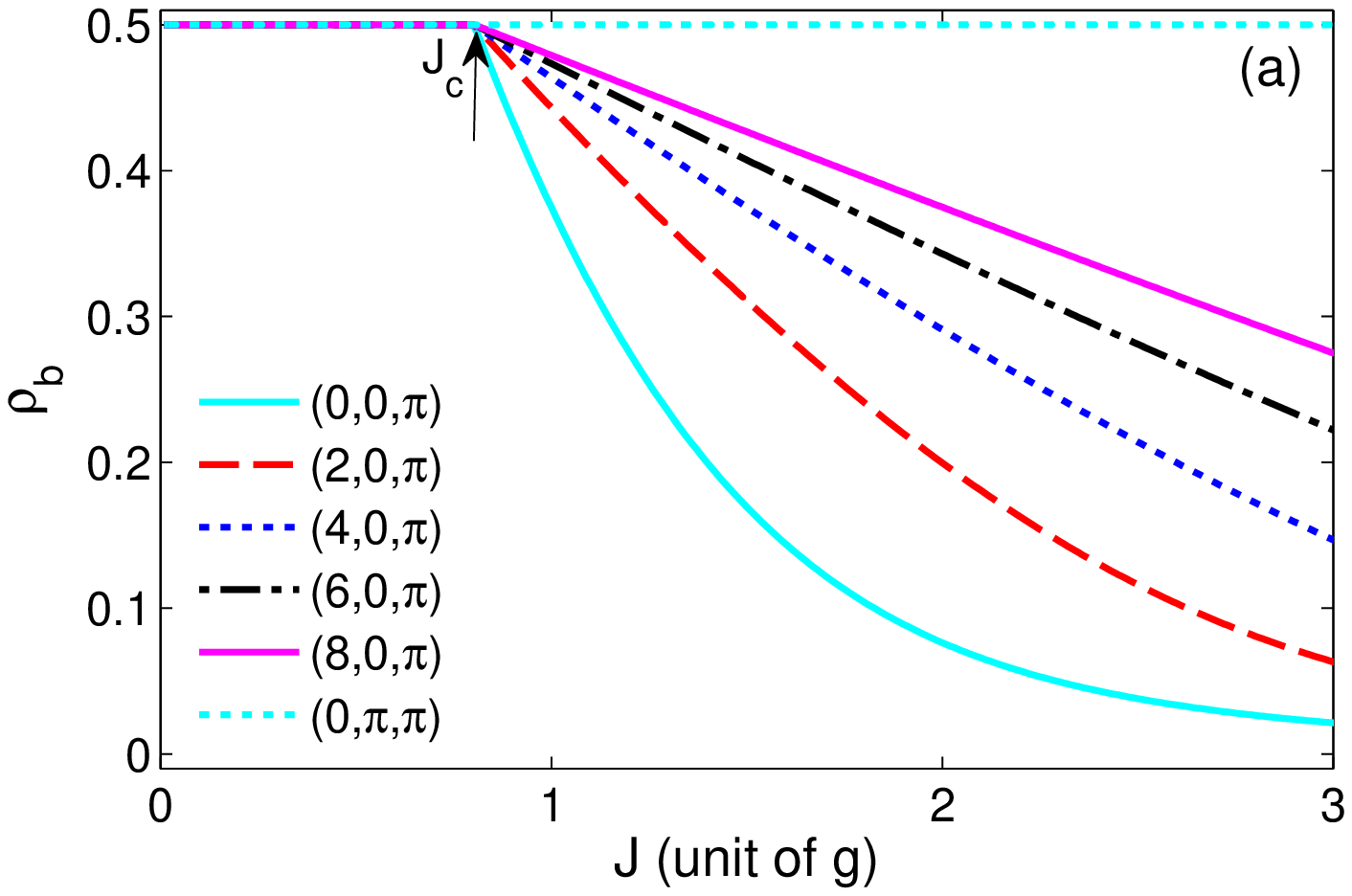}\\[3mm]
\includegraphics[width=70mm]{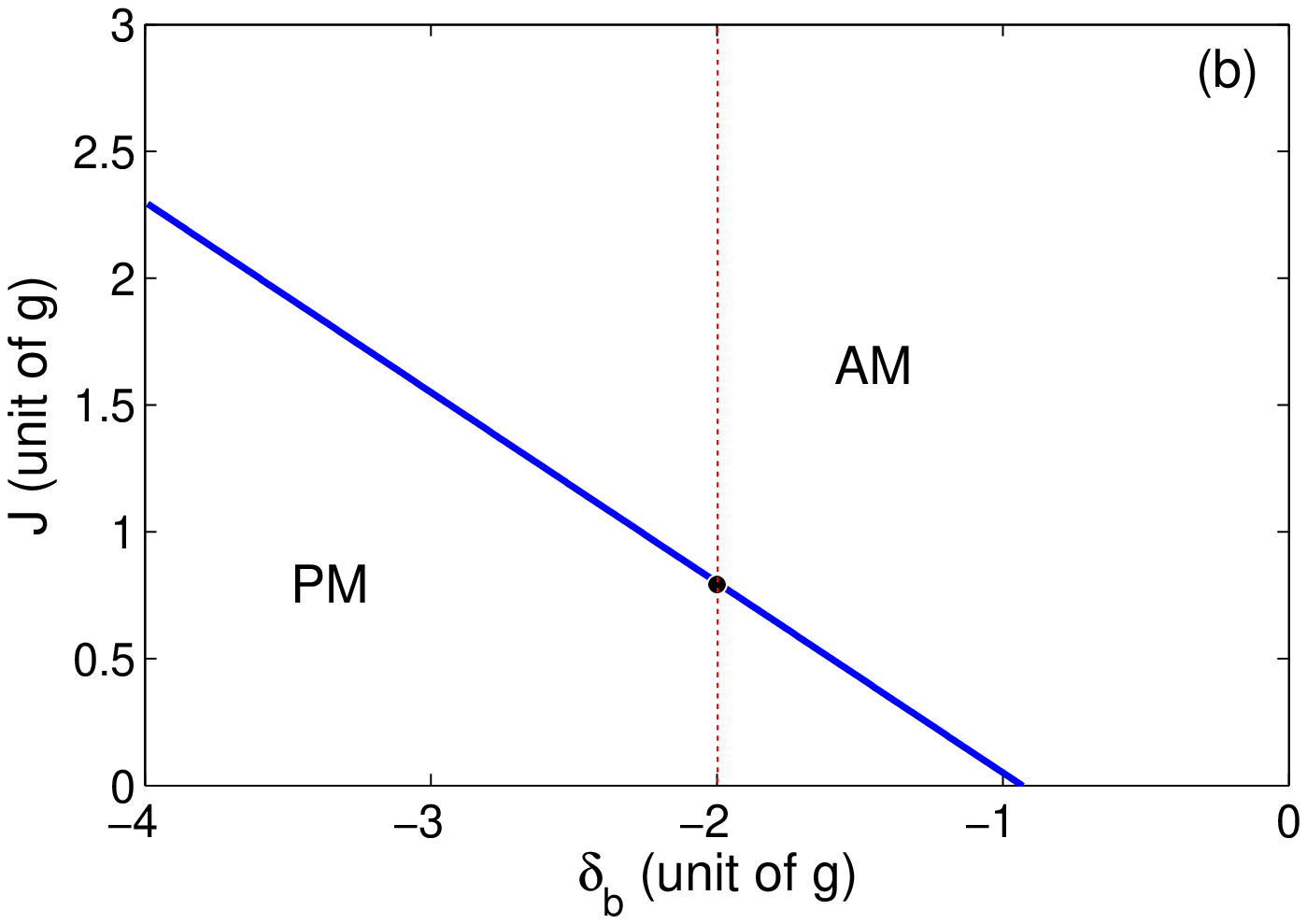}
\caption{(Color online) (a)  The density of molecules $\rho_b$ versus $J$ of the fixed points with $z=0$ for $\widetilde{U}=0$ marked with four cyan lines. The parameters are $g=1$ and $\delta_b=-2$. $\widetilde{U}$, $2\phi_a$ and $\phi$ are indicated in the legend. The ground states for nonzero atomic interaction $\widetilde{U}=2, 4, 6, 8$ are also shown. (b) The phase diagram of the ground state in MFA in the parameter space of $\delta_b$ and $J$ with unit of $g$. The pure molecule (PM) phase and the mixed atom-molecule (AM) phase are separated by a critical line (marked with blue solid line). We focus our thoughts on the red dotted line and the critical point ($(\delta_b=-2g, J_\texttt{C}=[3-\sqrt2]g/2)$ marked with a black filled circle).} \label{fig:fixed_point1}
\end{figure}

Since  the fixed points correspond to the eigenvalues of the nonlinear system \cite{FP_EV}, we can study the static properties of the system by solving the fixed points.   The fixed points  requires that  $\frac{d}{dt}\phi_{a}=0$, $\frac{d}{dt}\phi=0$, $\frac{d}{dt}z=0$, $\frac{d}{dt}\rho_{b}=0$, which can give many fixed point solutions.  In the following discussion, we only focus on the kind of  fixed points with $z=0$ since they contain the ground and the first excited state. Such a kind of fixed points read as,
\begin{eqnarray}
\label{fixed_points}
\rho_{b}&=&\left[\frac{\Delta\pm\sqrt{\Delta^2+6g^2}}{6g\cos\phi}\right]^2,\nonumber\\
z&=&0\quad\quad \phi=0 (\pi)\quad\quad 2\phi_a=0(\pi),
\end{eqnarray}
where  $\Delta=3\delta_b/2+2J\cos(2\phi_a)$, `$+$' is taken for  $\phi=0$, and  `$-$' is taken for  $\phi=\pi$.
The Equation~(\ref{fixed_points}) is obtained in the  assumption of  $\delta_{12}=0$ and $\widetilde{U}=U+U^\prime=0$. Note that for the case of $\widetilde{U}\neq0$, we can not give the analytical solution of the fixed point. Whereas,
we find that the existence of $\widetilde{U}$ does  not change the boundary between the pure molecule (PM) phase and the mixed atom-molecule (AM) phase.

In Fig.~\ref{fig:fixed_point1}(a), we plot the dependence of density of molecules on hopping strength for the ground and first excited states. In such a figure, the three quantities in the legend denote the values of $\widetilde{U}$, $2\phi_a$ and $\phi$.  The lines with $(\widetilde{U},0,\pi)$ correspond to the ground states for different $\widetilde{U}$, and the line with $(0,\pi,\pi)$  corresponds to the first excited state for $\widetilde{U}=0$.  Note that   the first excited state can not be influenced by the  value of $\widetilde{U}$,  so the line with $(0,\pi,\pi)$ can also  denote the first excited state for different interaction strength $\widetilde{U}$. From Fig.~\ref{fig:fixed_point1}(a), we can find that  the ground state is a pure molecule state when $J<J_\texttt{c}$, and it is a mixed atom-molecule state when $J>J_\texttt{c}$. Here the critical value $J_\texttt{c}=(3-\sqrt{2})g/2$ which is dependent on the energy detuning $\delta_b$ (here $\delta_b=-2g$) but not on the interaction strength $\widetilde{U}$.  Meanwhile, we give the general phase diagram of the ground state in the parameter space of $\delta_b$ and $J$  in Fig.~\ref{fig:fixed_point1} (b). The boundary between PM phase and AM phase is $3\delta_b/2+2J=-\sqrt2g$ with $g=1$ (corresponding to the blue solid line in Fig.~\ref{fig:fixed_point1} (b)).
Note that for the case of $J=0$, the critical point $\delta_b=-\sqrt{8/9}g)$ which agrees well with the result in ref.~\cite{QPT_Fu}.

\begin{figure}[tbph]
\includegraphics[width=80mm]{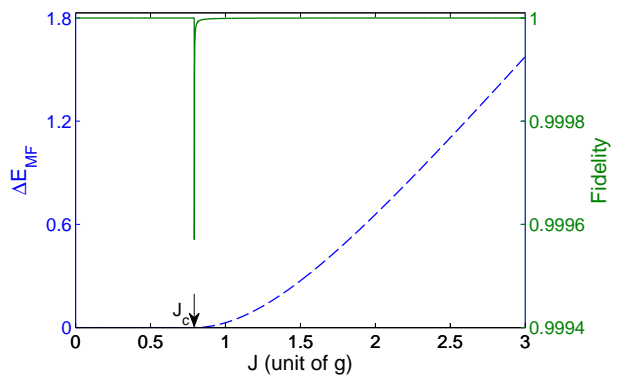}
\caption{(Color online)
Atomic hopping strength ($J$) profiles of the energy gap between the first excited state and the ground state ($\Delta E_{\mathrm{MF}}$) and of the fidelity of the ground state in MFA. The parameters are $\delta_b=-2g$ and $\delta J=0.001g$.} \label{fig:classical_analysis}
\end{figure}

In order to further confirm the phase boundary, we plot  profiles of the energy gap between the first excited state and the ground state ($\Delta E_{\mathrm{MF}}$) and of the fidelity of the ground state for a fixed energy detuning $\delta_b$ in MFA.  From Fig.~\ref{fig:classical_analysis}, we find that the energy gap between the ground and first excited state  is zero for $J<J_\texttt{c}$ and nonzero for $J>J_\texttt{c}$ for any interaction strength $\tilde{U}$. That implies that the transition from PM phase  to AM phase happens on the point where the energy degeneracy between ground and first excited state is lifted~\cite{T-QPT}. We know that the fidelity for eigenstates can also be used to characterize the phase transition. Here the fidelity for the ground state is defined as $F_{\texttt{MF}}=\langle\Psi_{\texttt{GS}}(J)|\Psi_{\texttt{GS}}(J+\delta J)\rangle$, where $|\Psi_{\texttt{GS}}(J)\rangle$ and $|\Psi_{\texttt{GS}}(J+\delta J)\rangle$ is the ground state of the system for different hopping strength  $J$ and $J+\delta J$ with $\delta J$ being  a very small quantity. From Fig.~\ref{fig:classical_analysis}, we also find that the fidelity has a sudden decrease from one at the critical point $J=J_\texttt{c}$, where the phase transition from a pure molecule phase to a mixed atom-molecule phase takes place. In a word, both the energy gap and the fidelity have fantastic phenomena at the critical point.

\section{Quantum phase transition}\label{sec:QPT}

\begin{figure}[tbph]
\includegraphics[width=80mm]{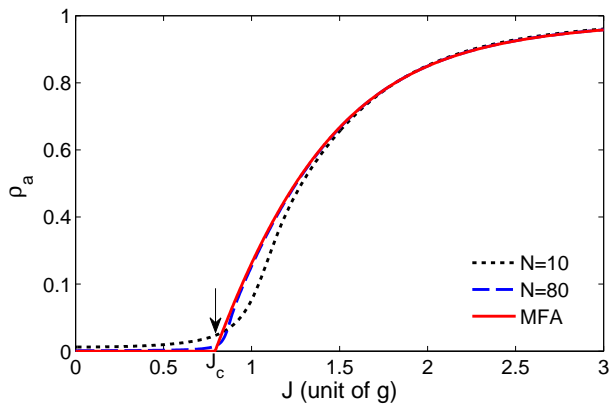}
\caption{(Color online) The total population of atoms ($\rho_a=1-2\rho_b$) in the ground state as a function of the hopping strength $J$  for different particle numbers. The red solid line is the MFA result (\ie~$N\rightarrow\infty$).} \label{fig:QPT_QC}
\end{figure}

To get insight into the QPT, we study the system in full quantum method. For a finite particle number $N$, where one molecule is counted as two particles, the Hamiltonian can be exactly diagonalized on the basis  of the Fock state $|n,N-n-2m,m\rangle$. Here  $n$ is the number of atoms for the first atomic component and  $m$ is the number of molecules. The dimension of the Fock basis is $(N/2+1)^2$. For convenience, the form of the Fock basis is signed  as  $|n,m\rangle$.
Then the eigenstate of the system can be written as \begin{equation}
|\Psi(J)\rangle=\mathop{\sum}_{n+2m\leq N}C_{n,m}(J)|n,m\rangle,
\end{equation}
where $C_{n,m}(J)$ are complex coefficients with parameter $J$.
The eigenvalues and the eigenstates of the system as a function of $J$ can be easily  obtained based on the method of exact diagonalization of Hamiltonian (\ref{model}).

In Fig.~\ref{fig:QPT_QC}, we plot  the total population of atoms for the ground state of the system with finite number of particles and in MFA
. From this figure, we can see that the QPT from a pure molecule state (the total population of the two atomic states is zero, \ie, $\rho_a=1-2\rho_b=0$) to a mixed atom-molecule state (the atomic population is nonzero, \ie, $\rho_a=1-2\rho_b>0$,) happens in the system as the atomic hopping strength reaches the critical value of  $J_N$. Note that the critical value of $J_N$ is very close to $J_\texttt{c}$ obtained in MFA, and $J_N\sim J_\texttt{c}$ in the limit of $N\rightarrow \infty$. In the following discussion, we will further characterize the quantum phase transition with the help of the concepts of energy gap, fidelity susceptibility and entanglement entropy.  Additionally, we will study the scaling behaviors of the system near the critical point $J_\texttt{C}$.

\subsection{Energy gap}
\begin{figure*}[tbph]
\includegraphics[width=55mm]{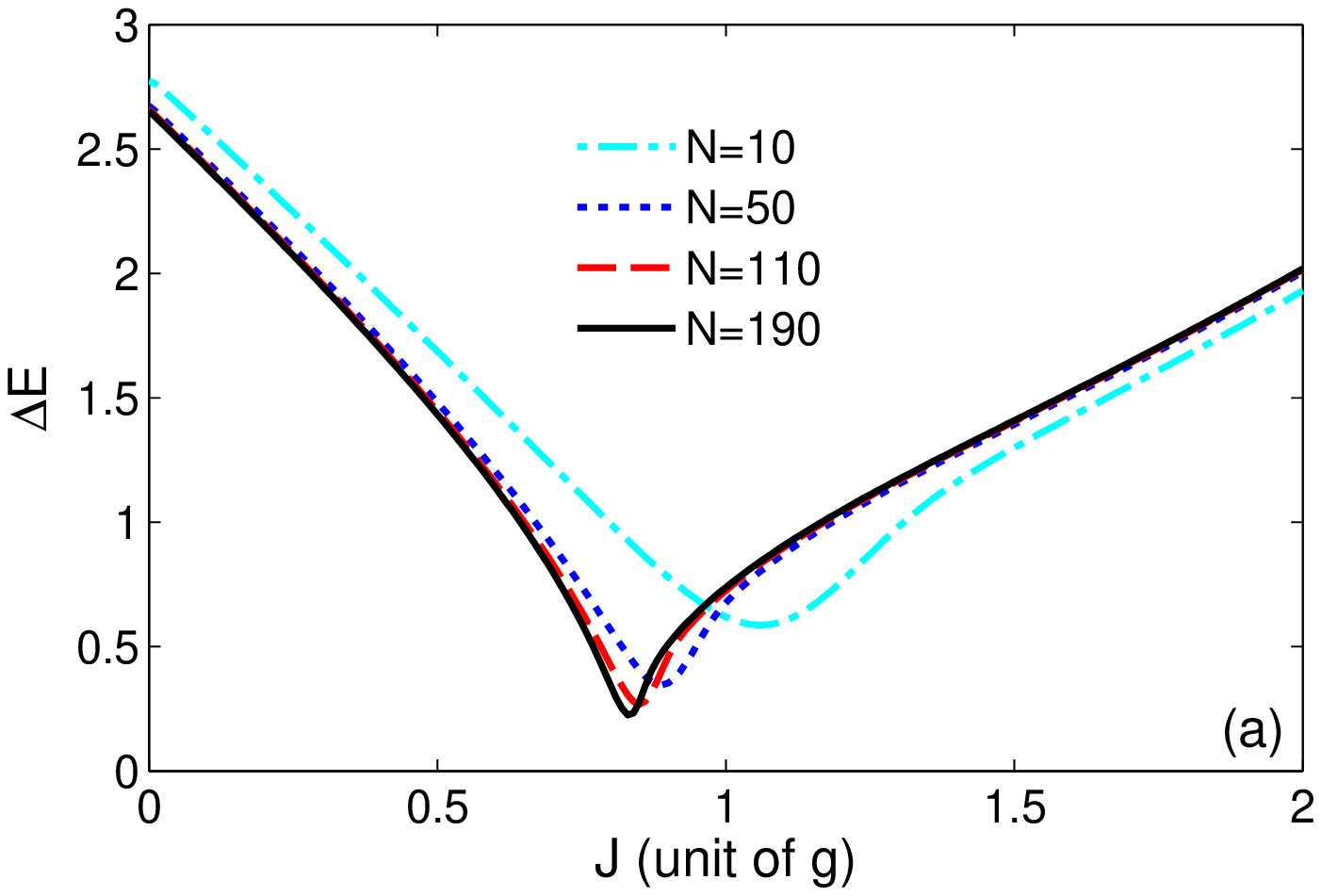}~~
\includegraphics[width=54mm]{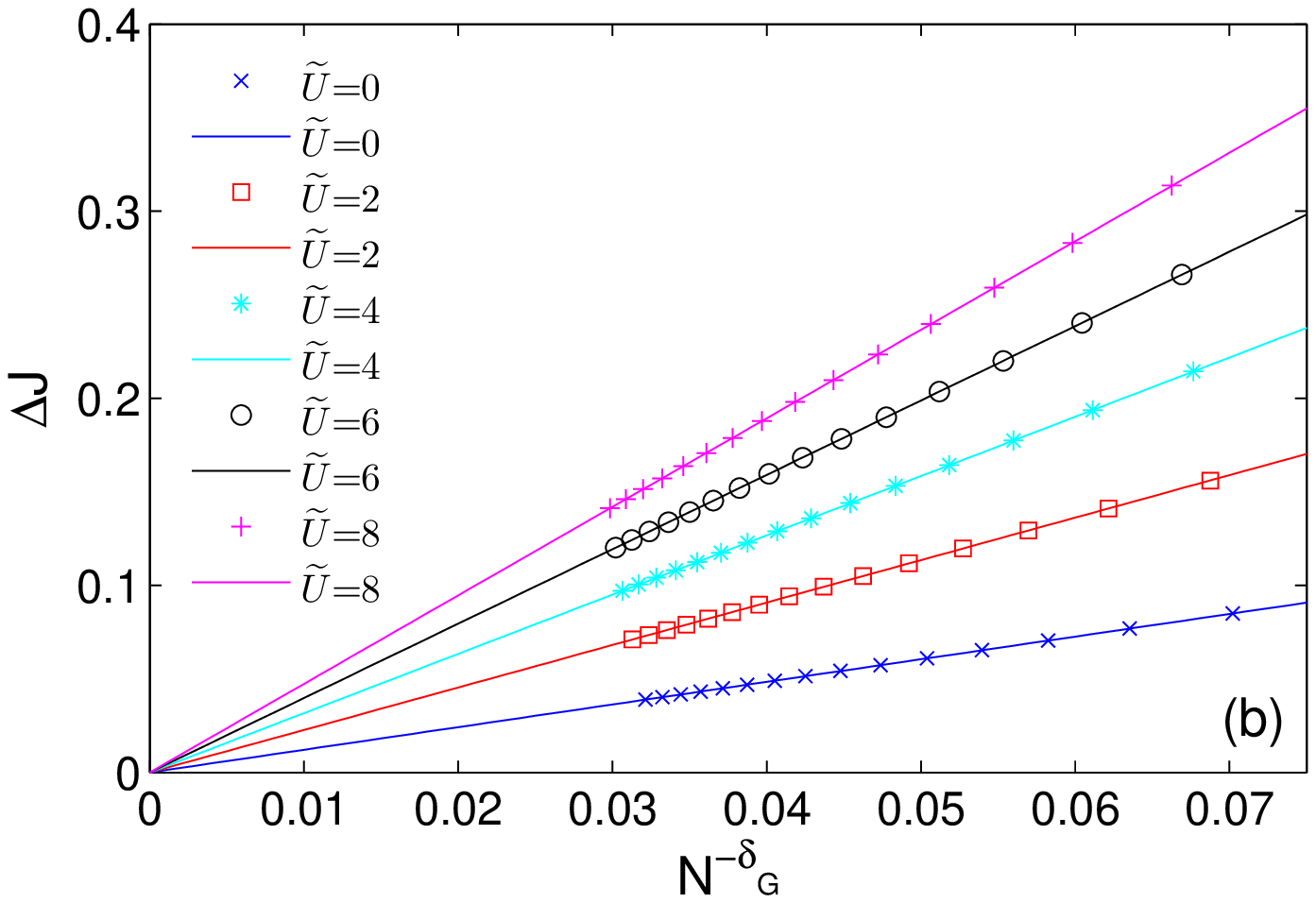}~~
\includegraphics[width=56mm]{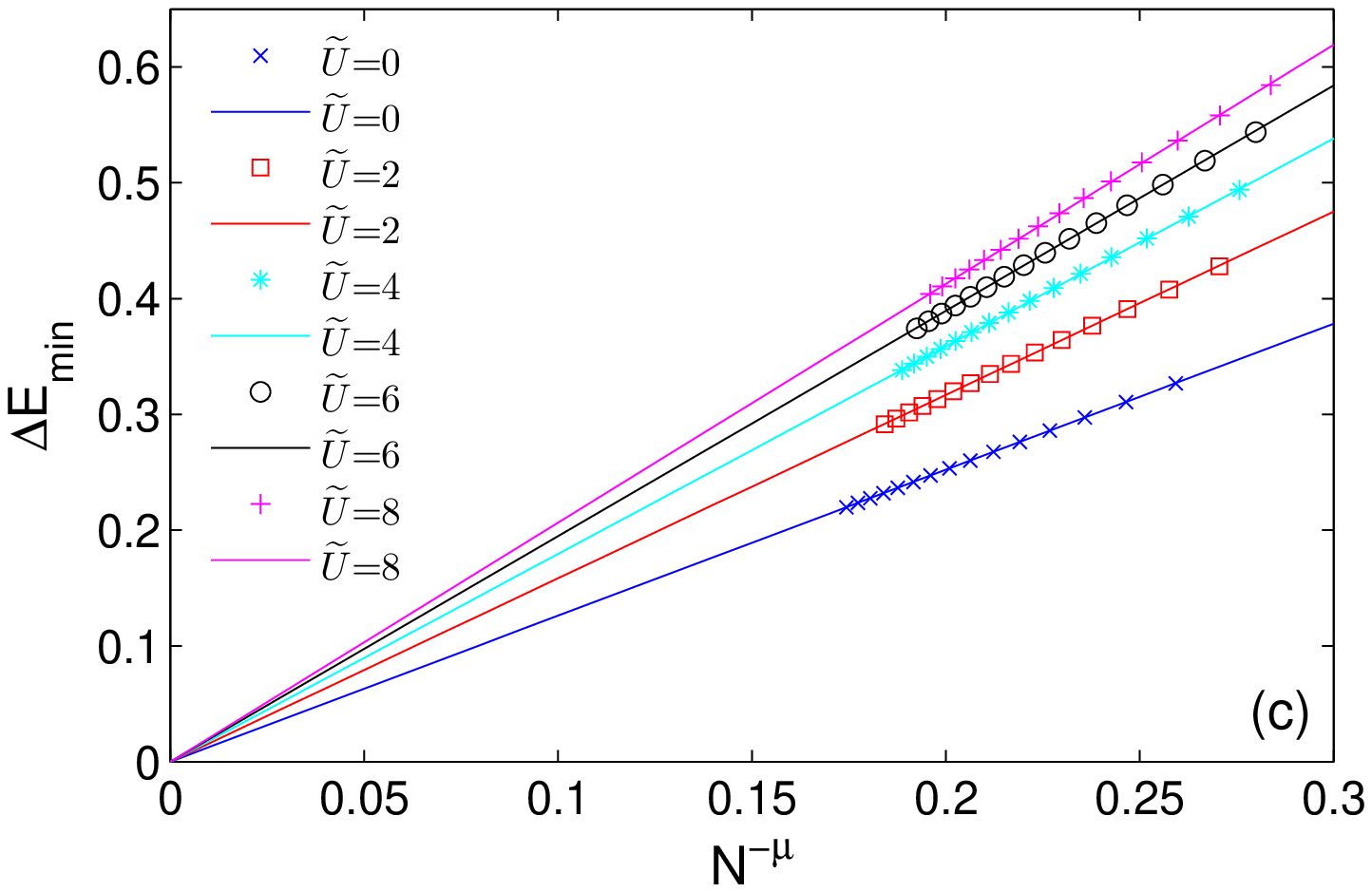}
\caption{(Color online) (a) Energy gap between the first excited state and the ground state versus the hopping strength $J$ for different particle numbers.
 (b) The finite-size scaling behaviors of $\Delta J=J_N-J_\texttt{C}$ for different  atomic interaction $\widetilde{U}$. $J_N$ is the parameter corresponding to $\Delta E_{min}$. (c) The finite-size scaling behaviors of $\Delta E_{min}$ for different atomic interaction  $\widetilde{U}$. The parameter is $\delta_b=-2g$.} \label{fig:gap_Q}
\end{figure*}

Now we are in the position to study  the QPT with the help of energy gap which has been used to characterize the QPT in ref.~\cite{AHM3_The, QPT_Fu, BA, QPT_AM2}. Here the energy gap is defined as the energy difference between the first excited state $E_1$ and the ground state $E_0$, \ie, $\Delta E=E_1-E_0$.
In Fig.~\ref{fig:gap_Q} (a),  the energy gap $\Delta E$ is shown as a function of $J$ for different particle numbers. For a fixed energy detuning $\delta_b$, the avoided level-crossing between the ground  and the first excited state appears near the critical hopping strength $J_\texttt{c}$ which is  the phase transition point given in mean-field approximation. From Fig.~\ref{fig:gap_Q} (a), we can see that for a finite particle number $N$, the energy gap reaches its minimum value $\Delta E_{min}$ at the critical points $J_N$ where the QPT from a pure molecule state to a mixed atom-molecule state happens.  We find that the critical hopping strength becomes closer with the increase of the particle number $N$.

To further characterize the finite-size effect present in Fig.~\ref{fig:gap_Q} (a), we show the scaling behaviors of $\Delta E_{min}$ and $\Delta J=J_N-J_\texttt{C}$ on the particle number $N$ for different interaction strength in Fig.~\ref{fig:gap_Q} (b) and (c).
In such two panels, the discrete points denotes the numerical results for finite particle numbers $N$, while the solid lines are the fitting functions which well show how the quantum results tend to MFA ones with the increase of particle numbers. From Fig.~\ref{fig:gap_Q} (b) and (c), we find that both $\Delta J$ and $(\Delta E)_{min}$ converge to zero  with different slopes for different
 $U$ when $N^{-\delta_G}$ and $N^{-\mu}$ approach zero. So that both $\Delta J$ and $(\Delta E)_{min}$  approach zero in the limit of $N\rightarrow\infty$, \ie, $\Delta E_{min}$ converges to zero and $J_N$ converges to $J_\texttt{C}$, which is agreed well with the results in MFA. Additionally, the critical exponents $\delta_G$ and $\mu$ for different $\widetilde{U}$ with slight difference are shown in Table \ref{tab:Gap}.

\begin{table}[!h] \caption{Finite-size scaling behaviors of the energy gap. Scaling exponents $\delta_G$ of $\Delta J\sim N^{-\delta_G}$ and $\mu$ of $\Delta E_{min}\sim N^{-\mu}$ at various $\widetilde{U}$ obtained by sampling system size in $N\in[60,200]$.}
\label{tab:Gap}
\tabcolsep 1.4mm
\par
\begin{center}
\begin{tabular}{cccccc}
\hline \smallskip $\widetilde{U}$ & $0$ & $2$ &$4$ & $6$ & $8$\\ \hline
\smallskip $\delta_G$ & $0.6488$ & $0.6539$& $0.6578$ & $0.6605$ & $0.6629$ \\
\smallskip $\mu$ & $0.3297$ & $0.3193$& $0.3147$ & $0.311$ & $0.3076$ \\
 \hline
\end{tabular}%
\end{center}
\end{table}

\subsection{Fidelity susceptibility}
\begin{figure*}[tbph]
\includegraphics[width=54mm]{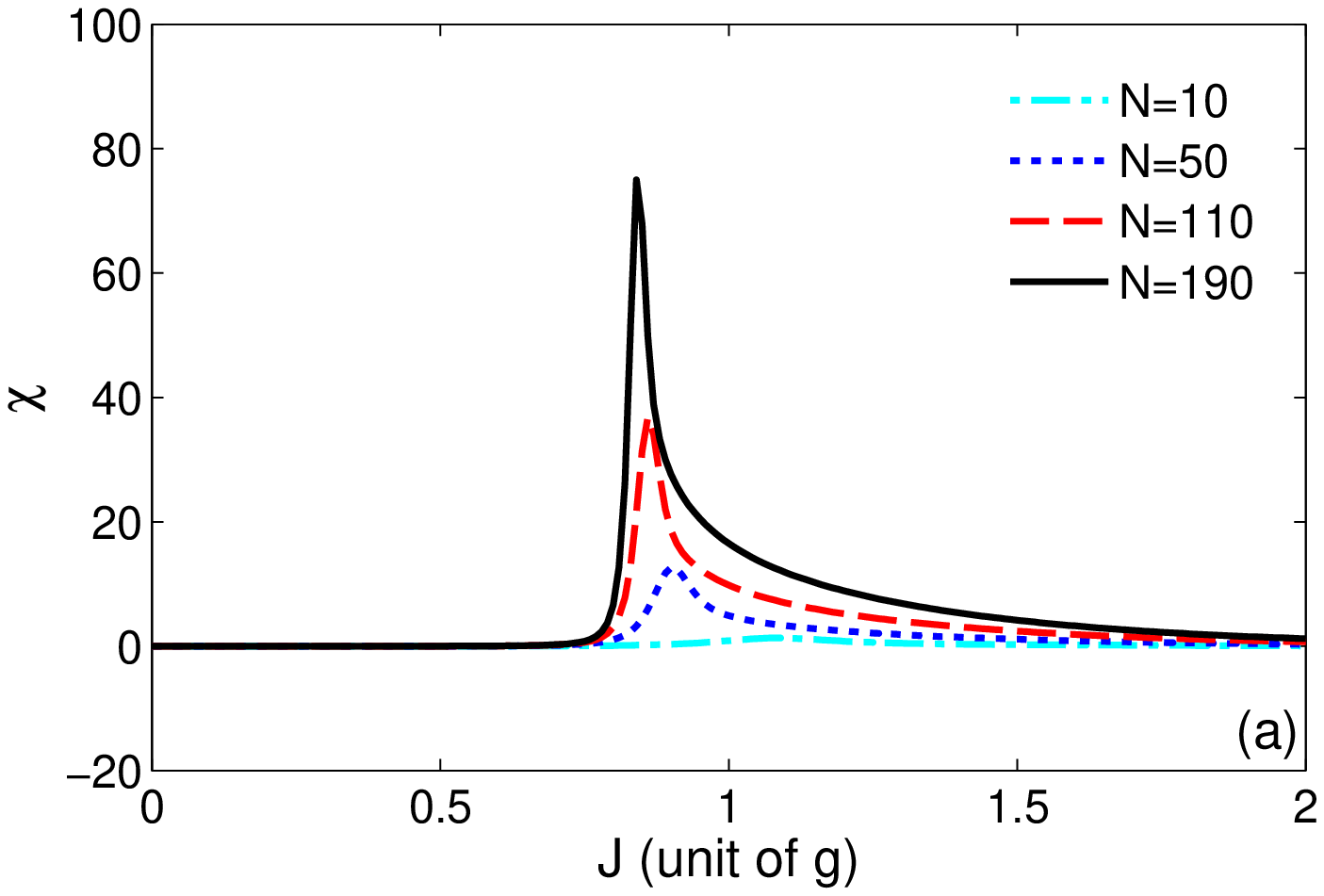}~~
\includegraphics[width=55mm]{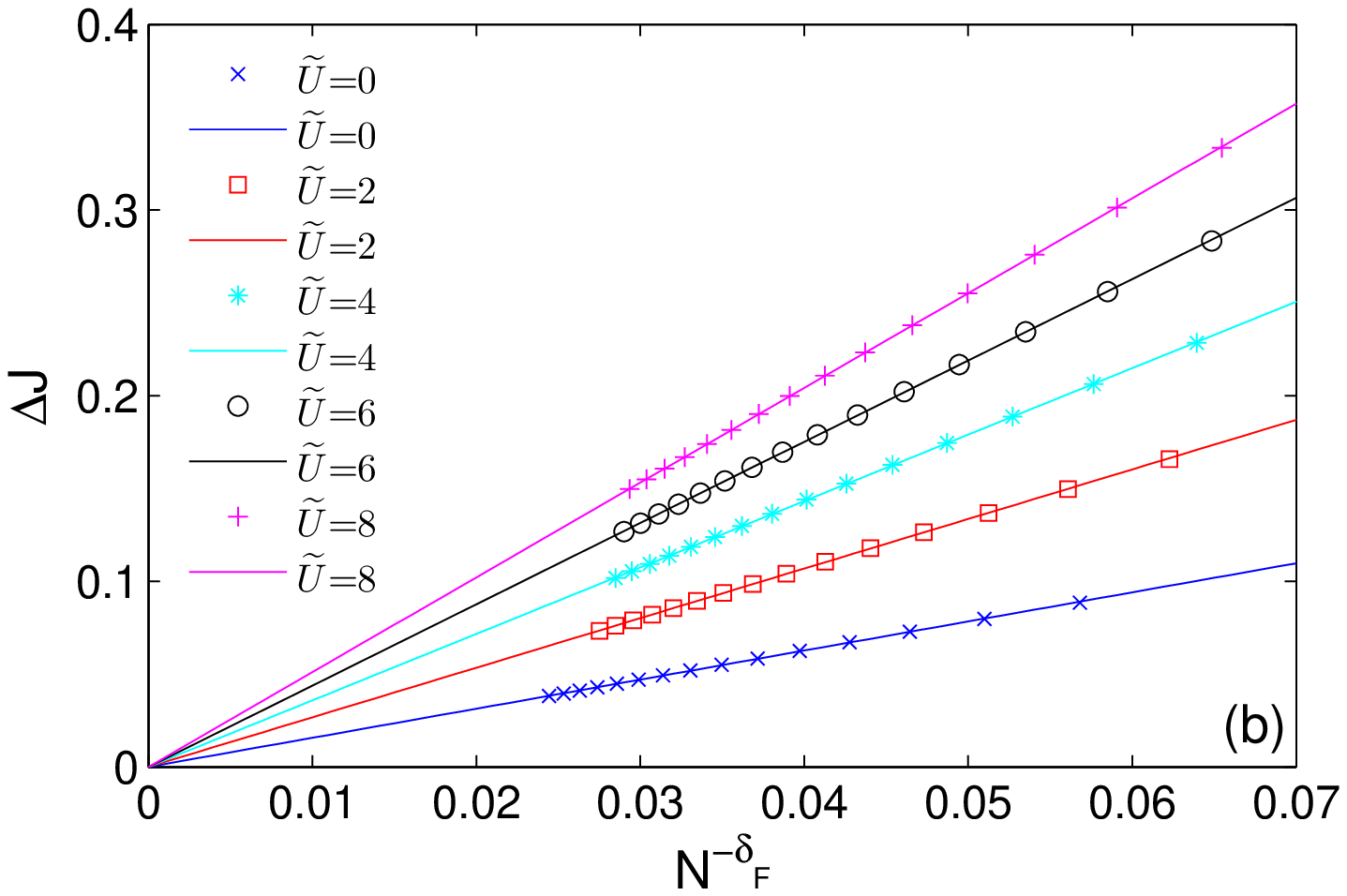}
\includegraphics[width=55mm]{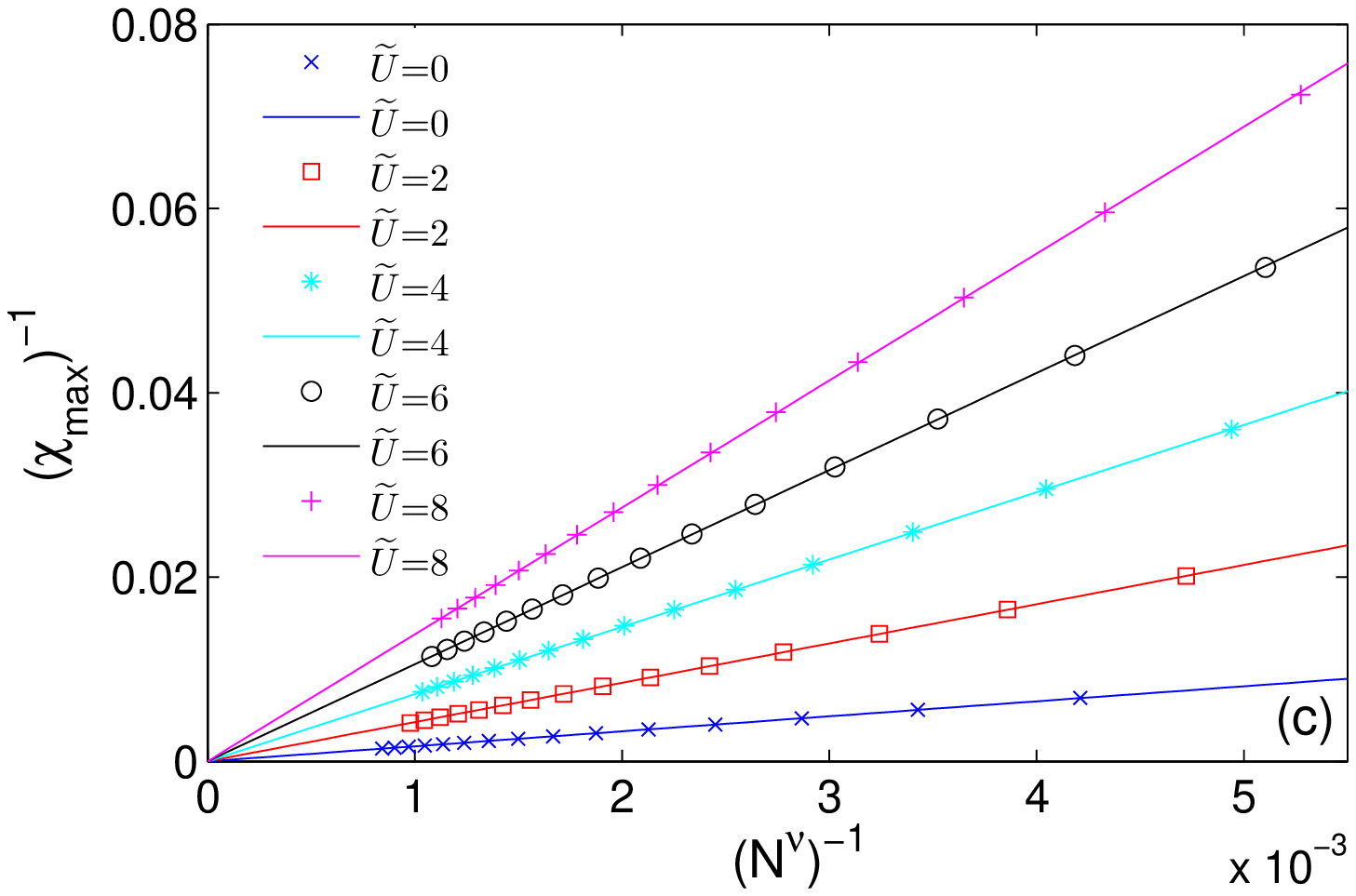}
\caption{(Color online) (a) The fidelity susceptibility for the  ground state as a function of the hopping strength $J$ for different particle numbers.
(b) The finite-size scaling behaviors of $\Delta J=J_N-J_\texttt{C}$ for different $\widetilde{U}$. $J_N$ is the parameter corresponding to $\chi_{max}$. (c) The finite-size scaling behaviors of $\chi_{max}$ for different $\widetilde{U}$. The parameter is $\delta_b=-2g$.
} \label{fig:fs_Q}
\end{figure*}

We know that the quantum fidelity can be used to characterize the quantum phase transition \cite{QPT_EE_DEE, fidelity, QPT_F_FS_EE, BA, QPT_AM2}, where the quantum fidelity is defined as the absolute value of the overlap between two ground states with an infinitesimal variation of the control parameter.
In order to study the effect of hopping strength $J$ on the QPT of ground state, the fidelity  can be written as
\begin{eqnarray}
\label{Fidelity}
F(J,\delta J)=|\langle\Psi_0(J)|\Psi_0(J+\delta J)\rangle|,
\end{eqnarray}
where $|\Psi_0(J)\rangle$ and $|\Psi_0(J+\delta J)\rangle$ are two ground states of the system with small parameter difference $\delta J$. We find that the value of fidelity is dependent on the value of $\delta J$ although there is a  sudden drop of the fidelity value near the critical point $J_\texttt{c}$ which is the phase transition point given in MFA.  In order to make up for this deficiency (\ie, the value of fidelity is dependent on the value of $\delta J$), we make use of the concept of fidelity susceptibility \cite{T-QPT, QPT_F_FS_EE, QPT_FS}.
In the first-order perturbation theory, we consider the hopping term as the perturbation term (\ie, $H=H_0+J H_J$). Then the fidelity susceptibility is defined as
\begin{eqnarray}
\label{Fs}
\chi(J)&=&\mathop{\lim}_{\delta\rightarrow0}\frac{2[1-F(J,\delta J)^2]}{(\delta J)^2}\nonumber\\
&=&\sum_{n\neq0}\frac{|\langle\Psi_{n}(J)|H_J|\Psi_{0}(J)\rangle|^2}{[E_n(J)-E_0(J)]^2},
\end{eqnarray}
which does not depend on the value of $\delta J$. It is also proved to be related to the correlation function \cite{fs_cf} which is used to show phase transition.

The numerical results of the fidelity susceptibility versus $J$ for $N=10, 50, 110, 190$ are shown in Fig.~\ref{fig:fs_Q} (a). We can see that
the fidelity susceptibility is about  zero when $J$ is far away from the critical point $J_N$,  while it  increases suddenly and reaches its maximum value at the critical point $J_N$ which is dependent on the particle number $N$. Meanwhile, we can
find that the maximum values of the fidelity susceptibility become larger and the  critical point $J_N$  become closer to $J_\texttt{C}$ as $N$ increasing. In order to study the effect of particle number on the critical behavior, we show the finite-size scaling behaviors of $\Delta J$ and $(\chi_{max})^{-1}$ for different $\widetilde{U}$ by power-law in Fig.~\ref{fig:es_Q} (b) and (c).

We can find clearly that both $\Delta J$ and $(\chi_{max})^{-1}$ converge to zero for various $\widetilde{U}$ with different slopes when $N^{-\delta_F}$ and $(N^{\nu})^{-1}$ approach zero, respectively.
In other words, $\Delta J$ is close to zero and $\chi_{max}$ is approximated to infinity when $N\rightarrow\infty$. We show the critical exponents $\delta_F$ and $\nu$ for different $\widetilde{U}$ without good convergency \cite{T-QPT, QPT_FS_Exponent}, in Table \ref{tab:Fs}.

\begin{table}[!h] \caption{Finite-size scaling behaviors of the fidelity susceptibility. Scaling exponents $\delta_F$ of $\Delta J\sim N^{-\delta_F}$ and $\nu$ of $\chi_{max}\sim N^{-\nu}$ at various $\widetilde{U}$ obtained by sampling system size in $N\in[60,200]$.}
\label{tab:Fs}
\tabcolsep 1.4mm
\par
\begin{center}
\begin{tabular}{cccccc}
\hline \smallskip $\widetilde{U}$ & $0$ & $2$ &$4$ & $6$ & $8$\\ \hline
\smallskip $\delta_F$ & $0.7005$ & $0.6781$& $0.6716$ & $0.6682$ & $0.6659$ \\
\smallskip $\nu$ & $1.336$ & $1.308$& $1.297$ & $1.289$ & $1.281$ \\
 \hline
\end{tabular}%
\end{center}
\end{table}

\subsection{Entanglement entropy}
\begin{figure*}[tbph]
~~~\includegraphics[width=62mm]{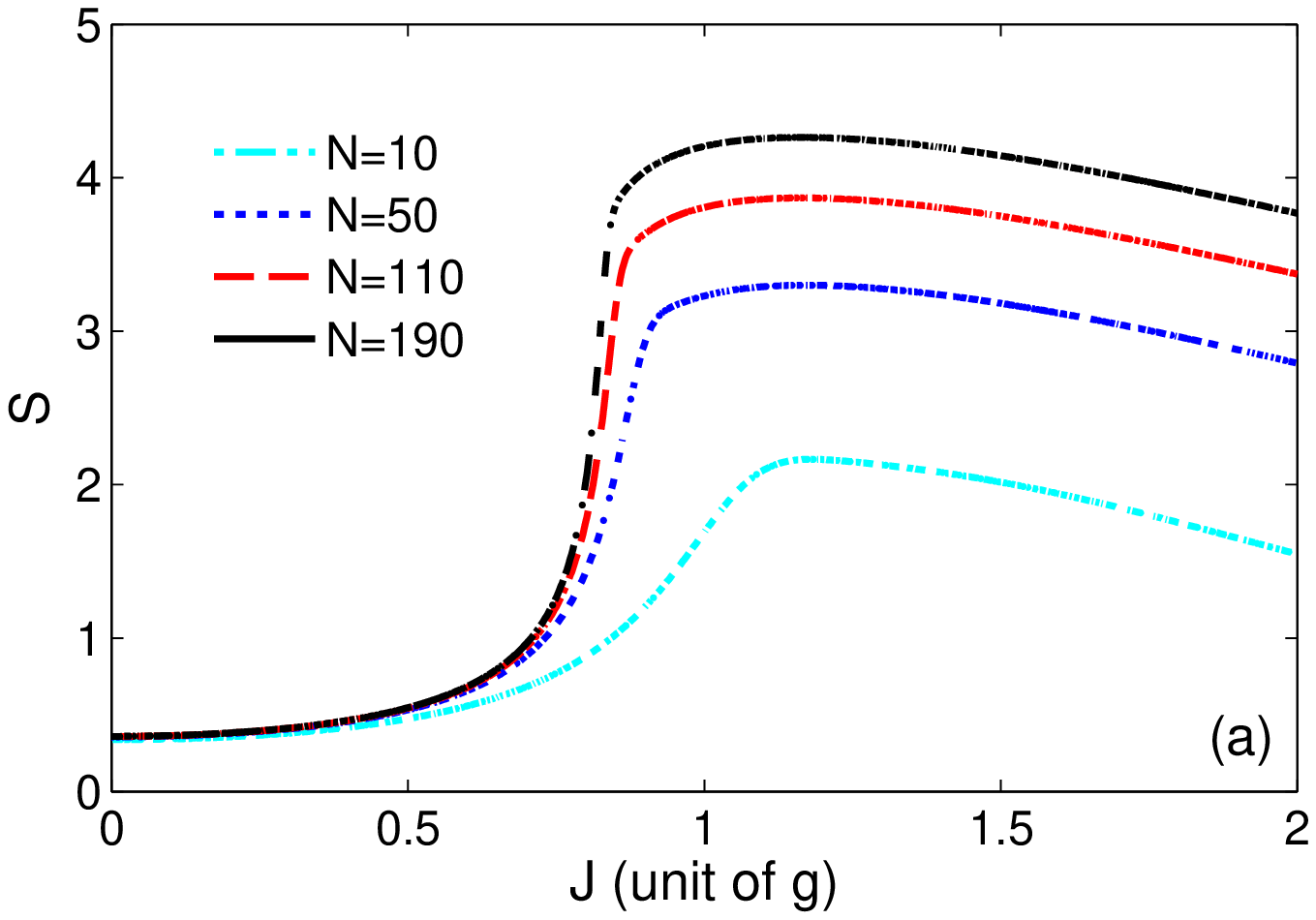}~~~~
\includegraphics[width=64mm]{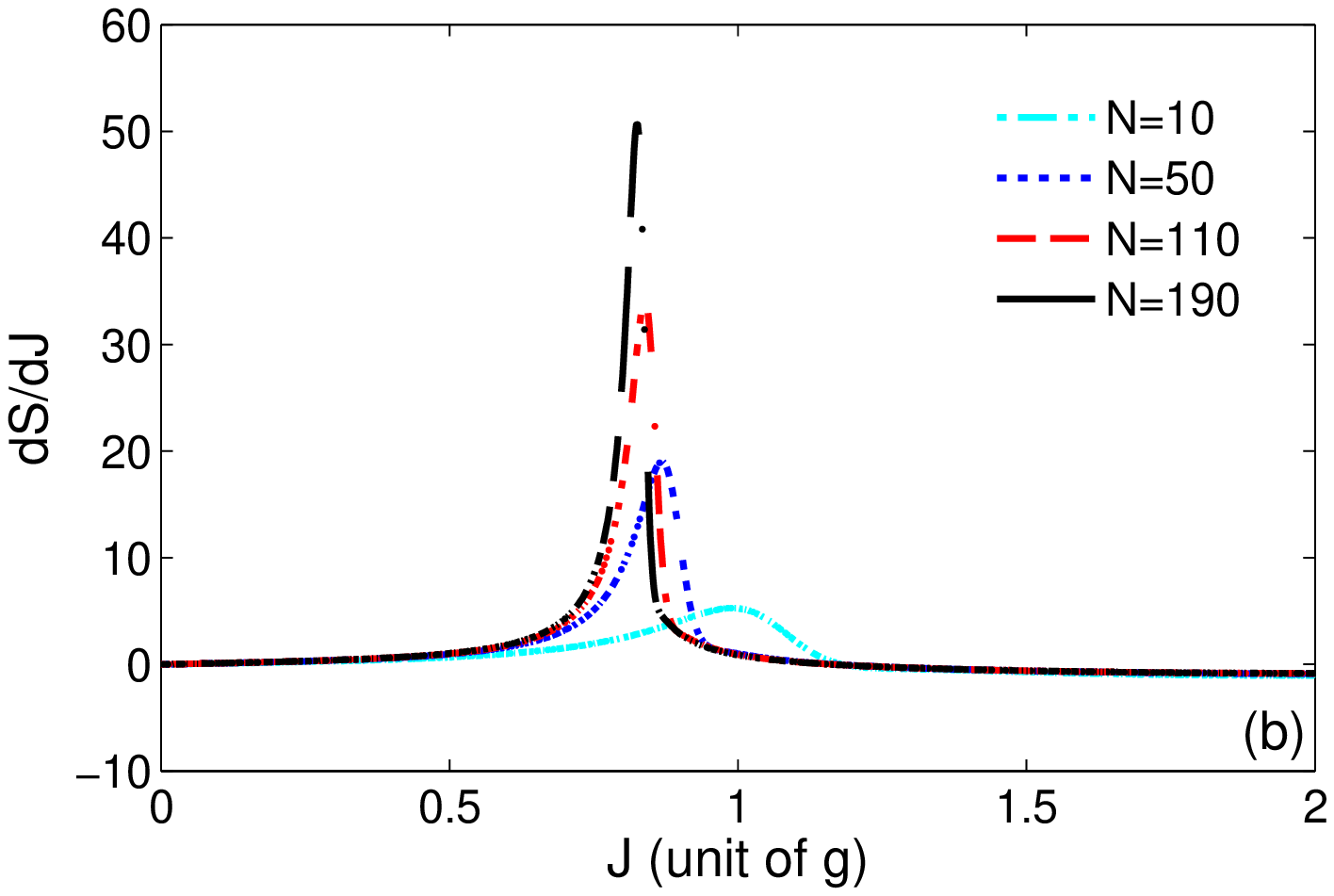}\\[2mm]
\includegraphics[width=65mm]{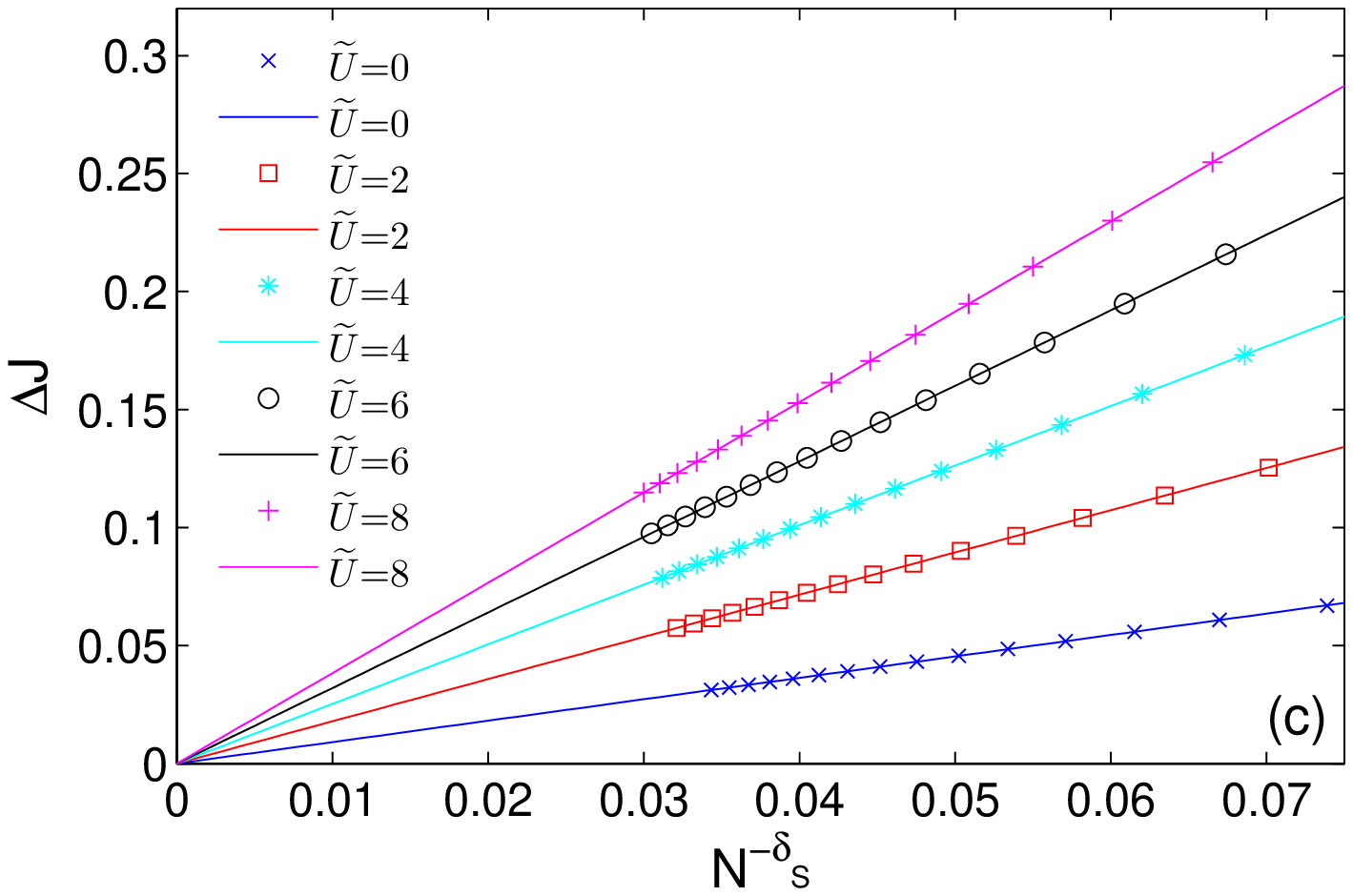}~~
\includegraphics[width=66mm]{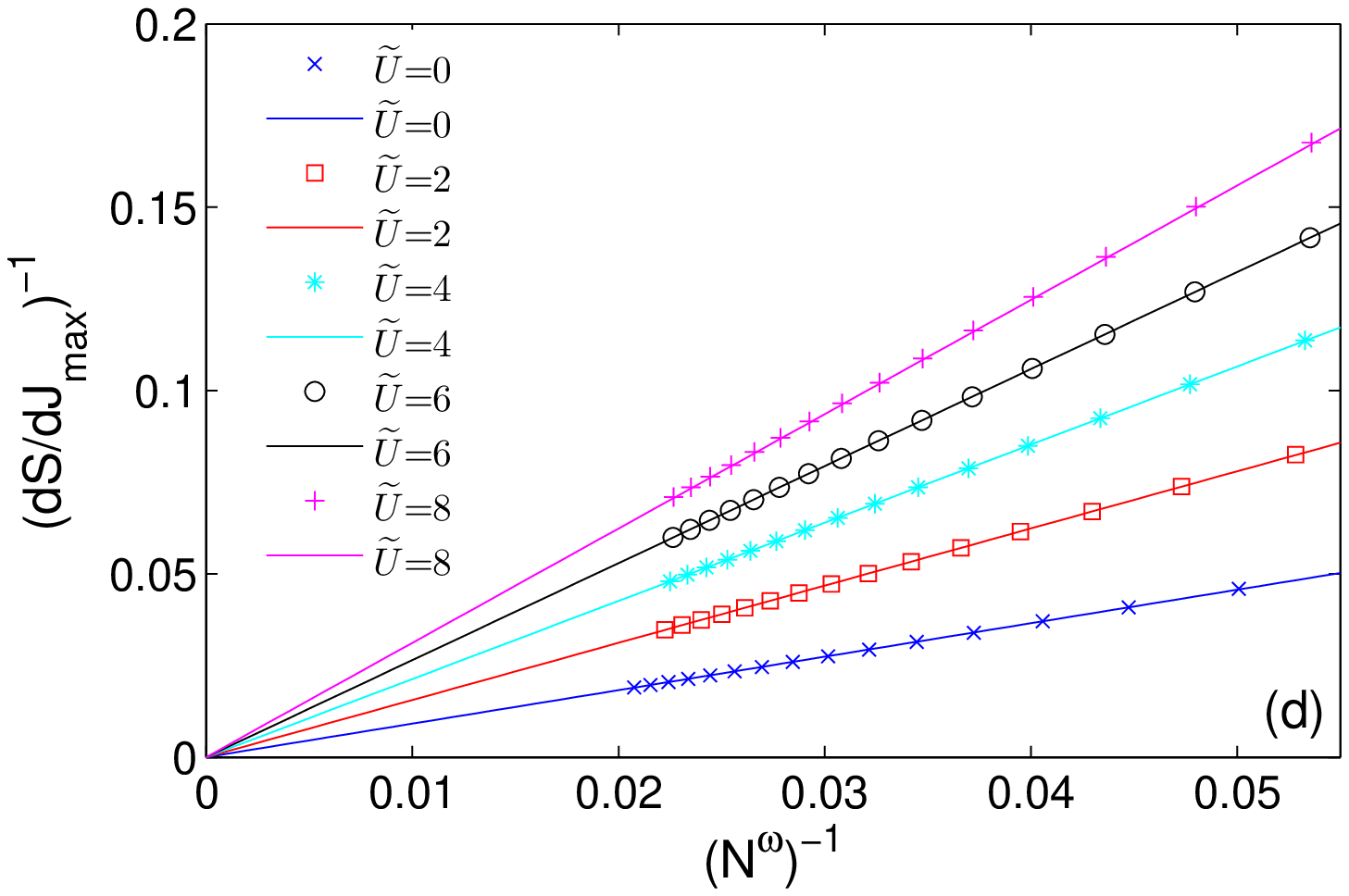}
\caption{(Color online) The entanglement entropy (a) and its first-order derivative (b) of the ground state  as a function of the hopping strength $J$ for different particle numbers. (c) The finite-size scaling behaviors of $\Delta J=J_N-J_\texttt{C}$ for different $\widetilde{U}$. $J_N$ is the parameter corresponding to $dS/dJ_{max}$. (d) The finite-size scaling behaviors of $dS/dJ_{max}$ for different  $\widetilde{U}$. The parameter is $\delta_b=-2g$.
} \label{fig:es_Q}
\end{figure*}

If the system can be viewed as a bipartite system, the entanglement entropy of its ground state is physically meaningful  and  has been used to characterize the quantum phase transition \cite{entanglement, Entanglement_QPT_Li, BA, QPT_F_FS_EE, QPT_AM2, QPT_EE}. The von Neumann entropy, one typical entanglement entropy, of a bipartite system $AB$ for a pure state $|\Psi\rangle$ is defined as
\begin{eqnarray}
S=-\texttt{Tr}_A(\rho_A\log_2\rho_A)=-\texttt{Tr}_B(\rho_B\log_2\rho_B),
\end{eqnarray}
where $\rho_{A(B)}=\texttt{Tr}_{B(A)}(|\Psi\rangle\langle\Psi|)$ is the reduced density matrix of the system with two subsystems A and B. In this paper, we consider the two atomic modes as subsystem A and the molecular mode as subsystem B.
From the Schmidt decomposition of pure state, we know that the entanglement entropies calculated from the reduced density operator of the subsystem A and B agree well with each other.

With the help of the definition of entanglement entropy, we plot the numerical results of the entanglement entropy between the two subsystems for the exact ground state in Fig.~\ref{fig:es_Q} (a) with different particle numbers $N=10, 50, 110, 190$.
In Fig.~\ref{fig:es_Q} (a), the maximum value of the entanglement entropy become larger as $N$ increasing.
Whereas, unlike the behavior of the transverse Ising model~\cite{TIM1, TIM2}, the entanglement entropy does not reach its maximum value  at the critical point $J_\texttt{C}$ even if $N\rightarrow\infty$. However, it is exciting to find that the sudden rise of the entanglement entropy takes place near the critical point $J_\texttt{C}=(3-\sqrt{2})/2$. In order to describe the quantum phase transition, the first-order derivative of the entanglement entropy with respect to $J$ is introduced. The numerical results of the first-order derivatives are shown in Fig.~\ref{fig:es_Q} (b). The maximum value $(dS/dJ)_{max}$ of first-order derivative of the entanglement entropy appears at the critical points $J_N$ which become closer to $J_\texttt{C}$ as $N$ increasing. So as to find the dependence of $(dS/dJ)_{max}$ and $\Delta J$ on the particle number $N$, we show their scaling behavior for different $\widetilde{U}$ in Fig.~\ref{fig:es_Q} (c) and (d).  We find that both $\Delta J$ and $(dS/dJ_{max})^{-1}$ converge to zero for various $\widetilde{U}$ with different slopes when $N^{-\delta_S}$ and $(N^{\omega})^{-1}$ approach zero. So that $\Delta J$ is close to zero and $dS/dJ_{max}$ is in limit of infinity when $N\rightarrow\infty$. Meanwhile , the critical exponents $\delta_S$ and $\omega$ for different $\widetilde{U}$ are shown in Table \ref{tab:DEE}.

\begin{table}[!h] \caption{Finite-size scaling behaviors of  the first-order derivative of the entanglement entropy. Scaling exponents $\delta_S$ of $\Delta J\sim N^{-\delta_S}$ and $\omega$ of $(dS/dJ)_{max}\sim N^{\omega}$ at various $\widetilde{U}$ obtained by sampling system size in $N\in[60,200]$.}
\label{tab:DEE}
\tabcolsep 1.4mm
\par
\begin{center}
\begin{tabular}{cccccc}
\hline \smallskip $\widetilde{U}$ & $0$ & $2$ &$4$ & $6$ & $8$\\ \hline
\smallskip $\delta_S$ & $0.6363$ & $0.649$& $0.6544$ & $0.6588$ & $0.6619$ \\
\smallskip $\omega$ & $0.7313$ & $0.7182$ & $0.7161$& $0.715$ & $0.714$ \\
 \hline
\end{tabular}%
\end{center}
\end{table}

\section{Summary}\label{sec:summary}
In this paper, we have studied the phase transition in an atom-molecule conversion system where the atoms can jump between two atomic hyperfine states. In mean-field approximation, we have given the the phase diagram for  the ground state, and shown that the phase boundary between  pure molecule phase and  mixed atom-molecule one  is only dependent on  the hopping strength $J$ and energy detuning $\delta_b$ but not dependent on the atomic interaction $\widetilde{U}$.   With the help of fixed points, we have studied the fidelity for the ground state and the energy gap between the first-excited  state and the ground one. We have found that the ground state changes from degeneracy to non-degeneracy and the fidelity decreases suddenly at the phase boundary $J_\texttt{C}=-\frac{3}{4}\delta_b-\frac{\sqrt{2}}{2}g$ with $g=1$, which implies that the energy gap and the fidelity can well characterize that phase transition.

In comparison to mean-field approximation, we have investigate the quantum phase transition of the system in full quantum method. Taking the total population of atoms as the order parameter, we have shown that the QPT from a pure molecule phase to a mixed atom-molecule phase happens at the critical point $J_N$ for different $N$ with a fixed energy detuning $\delta_b$. Note that the value of $J_N$ depends on both the value of $\delta_b$ and the particle number $N$ of the system. We have further characterized the QPT with the help of the energy gap between the first-excited state and the ground state, the fidelity susceptibility of the exact ground state, and the entanglement entropy and its first-order derivative between the atomic subsystem and molecular subsystem, respectively, which give the same critical point $J_N$.  We have also  shown that the critical point $J_N$ approaches to $J_\texttt{C}$ with the increase of particle number $N$ through studying the  finite-size scaling behaviors of the energy gap, fidelity susceptibility and first-order derivative of entanglement entropy. So that in the limit of $N\rightarrow\infty$, the MFA and full quantum methods can give the same phase boundary.
Our results enrich the phenomena of QPT in multiple component system especially atom-molecule conversion system, and indicate that one can control the QPT of  the  atom-molecule conversion  system by changing the hopping strength between the two atomic hyperfine states  as well as the  atom-molecule energy detuning.

\section*{Acknowledgments}

The work is supported by NSFC Grant No. l10674117, No. 11074216, No. 11104244, Zhejiang NSFC Grant No.Y4110063, and partially by PCSIRT Grant No. IRT0754.

\end{document}